\documentclass[12pt]{article}
\usepackage{amssymb}
\usepackage{amsmath}
\usepackage{amstext}
\usepackage{graphicx,epsfig}
\usepackage{epsfig}
\usepackage{verbatim} 
\usepackage{fancyhdr}
\usepackage{fancybox}
\usepackage{color}
\usepackage{ulem,bbold}
\usepackage{enumitem}
\usepackage{subfigure}
\usepackage{bbm}
\usepackage{parskip}
\newcommand{\Comment}[1]{{}}
\definecolor{MyDarkBlue}{rgb}{0.15,0.15,0.45}
\usepackage[linktocpage=true]{hyperref}
\hypersetup{
colorlinks=true,
citecolor=MyDarkBlue,
linkcolor=MyDarkBlue,
urlcolor=MyDarkBlue,
}

\newcommand{\be}{\begin{equation}}  
\newcommand{\ee}{\end{equation}}  
\newcommand{\bea}{\begin{eqnarray}}  
\newcommand{\eea}{\end{eqnarray}}
\newcommand{\nn}{\nonumber}

\newcommand{\la}{\langle}
\newcommand{\ra}{\rangle}
\newcommand{\half}{\frac{1}{2}}

\def\({\left(}
\def\){\right)}

\def\L{{\cal L}}

\def\mpl{M_{\rm Pl}}
\def\p{\partial}
\def\mn{_{\mu \nu}}

\numberwithin{equation}{section}

\begin{document}


\begin{flushright}
{UCSD-PTH-12-11}
\end{flushright}

\begin{center}
{\Large \bf{A Covariant Master Theory for Novel Galilean Invariant Models and Massive Gravity}}
\end{center} 
 \vspace{.2truecm}
\thispagestyle{empty} \centerline{
{\large  { Gregory Gabadadze${}^{a,b,}$}}\footnote{E-mail address: \Comment{\href{mailto:: gg32@nyu.edu}}{\tt gg32@nyu.edu}} 
{\large  { Kurt Hinterbichler${}^{c,}$}}\footnote{E-mail address: \Comment{\href{mailto:kurthi@physics.upenn.edu}}{\tt kurthi@physics.upenn.edu}} 
{\large  { Justin Khoury ${}^{c,}$}}\footnote{E-mail address: \Comment{\href{mailto:jkhoury@sas.upenn.edu}}{\tt jkhoury@sas.upenn.edu}}}
\centerline{
{\large  { David Pirtskhalava${}^{d,}$}}\footnote{E-mail address: \Comment{\href{mailto:pirtskhalava@physics.ucsd.edu}}{\tt pirtskhalava@physics.ucsd.edu}}
{\large  { Mark Trodden ${}^{c,}$}}\footnote{E-mail address: \Comment{\href{mailto:trodden@physics.upenn.edu}}{\tt trodden@physics.upenn.edu}}
                                                          }


\centerline{\it $^a$Center for Cosmology and Particle Physics,
Department of Physics,}
\centerline{\it New York University, New York,
NY, 10003, USA}


\centerline{\it $^b$ Perimeter Institute for Theoretical Physics,}
\centerline{\it 31 Caroline St. N, Waterloo, Ontario, Canada, N2L 2Y5}


\centerline{{\it ${}^c$ 
Center for Particle Cosmology, Department of Physics and Astronomy,}}
 \centerline{{\it University of Pennsylvania, Philadelphia, PA 19104, USA}} 
 

\centerline{\it $^d$Department of Physics, University of California, San Diego, La Jolla, CA 92093 USA}

\begin{abstract}
Coupling the galileons to a curved background has been a tradeoff between maintaining second order 
equations of motion, maintaining the galilean shift symmetries, and allowing the background metric 
to be dynamical. We propose a construction which can achieve all three for a novel class of galilean invariant 
models,  by coupling a scalar with the galilean symmetry to a massive graviton.  
This generalizes the brane construction for galileons, by adding to the brane a 
dynamical metric, (non-universally) interacting with the galileon field.  Alternatively, it can 
be thought of as an extension of the ghost-free massive gravity, or as a massive graviton-galileon 
scalar-tensor theory.  In the decoupling limit of these theories, new kinds of galileon invariant 
interactions arise between the scalar and the longitudinal mode of the graviton.  
These have higher order equations of motion and infinite powers of the field, yet are ghost-free.
\end{abstract}

\newpage

\thispagestyle{empty}
\newpage
\setcounter{page}{1}
\setcounter{footnote}{0}

\section{Introduction}
\parskip=5pt
\normalsize

The {\it Galileons}~\cite{Luty:2003vm,Nicolis:2008in} are higher derivative scalar field theories with many interesting and important properties, including second order equations of motion and novel non-linearly realized shift symmetries.  Originally formulated in flat space, it is not straightforward to couple the galileons to a curved background.  Implementing a universal coupling by a na\'ive replacement of partial derivatives by covariant derivatives results in a theory with higher order equations of motion for the metric.  It is possible to add non-minimal couplings to restore the second order equations of motion~\cite{Deffayet:2009mn,Deffayet:2009wt}, however, doing so inevitably results in terms that violate the galileon shift symmetries.

The brane construction of~\cite{deRham:2010eu} is an illuminating way to build the galileons\footnote{For a complementary construction of these theories as Wess-Zumino terms, see~\cite{Goon:2012dy}.}.  One imagines that our space-time is a 3-brane floating in some higher dimensional bulk spacetime, and the brane-bending modes then become the galileons.  By extending this construction to curved bulks and branes, it becomes possible to couple the galileons to different background geometries while preserving generalized versions of the galileon shift symmetries, which are now associated with isometries of the bulk~\cite{Burrage:2011bt,Goon:2011qf,Goon:2011uw,Goon:2011xf}.  The second order equations of motion are also preserved in this approach.  However, the metric is a fixed background and is not dynamical.  Making it dynamical corresponds to turning on a zero mode for the bulk metric, which breaks the isometries and hence the galileon symmetry~\cite{Goon:2012mu}.

On a parallel front, in recent years it has become possible to construct ghost-free (de Rham, Gabadadze, Tolley (dRGT)) theories of massive gravity~\cite{deRham:2010ik,deRham:2010kj} (See~\cite{Hinterbichler:2011tt} 
for a theory review, and \cite {Goldhaber:2008xy} for phenomenology review). These theories can be interpreted as the theory of a 3-brane embedded in a 3+1 dimensional bulk (i.e. a spacetime-filling embedding), in which a dynamical metric is put on the brane~\cite{ArkaniHamed:2002sp}, and the brane worldvolume action takes the form  given 
in \cite {deRham:2010kj}. The brane-bending modes become pure gauge modes, or St\"uckelberg fields, which, in the presence of interaction terms mixing the dynamical bulk metric with the induced metric, give the graviton a mass.
Interestingly, the Galileon terms emerge in ghost-free massive gravity in the decoupling 
limit \cite {deRham:2010gu,deRham:2010ik}.

In this paper, we combine elements of the brane construction for galileons and massive gravity to yield a novel 
theory that couples a multiplet of scalar fields $\pi^I$ (where $I$ is a flavor index, running from $1$ to $N$)  to a metric $g_{\mu\nu}$ in a way that possesses all three desirable features: no extra propagating degrees of freedom, a galileon symmetry, and dynamics for the metric. We should stress that what we mean by a ``galileon" in 
this work is a generic scalar field $\pi^I$, nontrivially transforming under the field-space galilean invariance of the theory
\be
\label{gt}
\pi^I\to \pi^I+\omega^I_\mu x^\mu ~,
\ee   
and propagating no extra degrees of freedom than those of a free field.
In particular, $\pi$ does not necessarily have to interact with itself or other  fields through the five standard galileon terms \cite{Nicolis:2008in} (or their multi-field generalizations \cite{Deffayet:2010zh,Padilla:2010de,Hinterbichler:2010xn}). Moreover, as we show below, a certain high-energy (``decoupling") limit of the theory is described by peculiar scalar interactions, which significantly differ from the standard galileon interactions, and yet their defining properties --- ghost freedom and galileon symmetry --- are retained.

In light of findings of Refs. \cite{Deffayet:2009mn,Deffayet:2009wt}, it is not surprising that a theory that can simultaneously achieve these properties is characterized by very special, non-universal couplings of the scalars $\pi^I$ to the dynamical metric. In particular, as already noted above, the absence of higher time derivatives in the equations of motion (or, equivalently, extra propagating ghost degrees of freedom) in theories with $\pi^I$ coupled to a massless graviton inevitably results in the breaking of galileon symmetries of the flat space theory. Based on the requirements of galilean invariance (which arises from nonlinearly realized broken higher-dimensional Poincar\'e symmetries and is therefore automatic in brane constructions), and ghost-freedom, we argue below that the metric that can naturally couple to the scalars $\pi^I$ describes a massive spin-2 field. The essential degrees of freedom on top of the $N$ scalars $\pi^I$ in our construction therefore are the five polarizations of a ghost-free massive graviton. 

Formally one can take different points of view towards the theory discussed below. On one hand, one can view it as a certain generalization of ghost-free massive gravity, consistently interacting with a set of scalars $\pi^I$ with the field-space galilean invariance \eqref{gt}\footnote{Another generalization of dRGT massive gravity by the ``quasi-dilaton", a scalar realizing a new global symmetry in the theory, has recently been considered in \cite{D'Amico:2012zv}. }. Alternatively, one can imagine a four-dimensional effective field theory, obtained by extending the brane construction of the galileons to allow for an additional intrinsic metric, describing dynamical gravity. Starting with the two non-interacting sectors, consisting of (a) the galileons (obtained via invariants formed from the induced metric in the standard way) and (b) dynamical gravity, one can ask whether it is possible to construct well-defined mixings/interactions between these sectors, which would not lead to extra propagating degrees of freedom. As we argue below, one is quite uniquely led to a generalization of the dRGT theories. The byproduct of this construction is that the dynamical metric on the brane will inevitably describe a massive graviton, nontrivially transforming under the galileon symmetries.

Interestingly enough, although governed by similar global symmetries, not all interactions of $\pi^I$ with the longitudinal scalar mode of the graviton in this theory fall into the categories of standard galileons or multi-galileons found in \cite{Nicolis:2008in,Deffayet:2010zh,Padilla:2010de,Hinterbichler:2010xn}. Unlike the usual galileon terms and their multi-field generalizations (of which there are a finite number for a finite number of fields), these interactions consist of an infinite series of terms with higher derivatives, and lead to higher order equations.  Yet, these theories are degenerate in the higher derivatives, and only two pieces of initial data per field are required to set the dynamics of the system, so in fact the theory is still free of extra propagating (potentially ghostly) degrees of freedom, albeit in a way that differs from the standard galileon interactions (which have purely second order equations).

The paper is organized as follows.  We start with reviewing the brane construction of galileons in section \ref{sec2}. Sections \ref{sec3} and \ref{sec4} describe the ghost-free actions for the induced and internal metrics that one can write within this framework, including a generalization of the recently proposed zero-derivative interactions, characterizing ghost-free massive gravity. In section \ref{sec5}, we introduce the basic model with a flat bulk and comment on its symmetries, while sections \ref{sec6} and \ref{sec7} deal with the decoupling limit of the simplest version of the theory and show how, at least in this limit, non-propagation of extra degrees of freedom is achieved despite the presence of higher derivative interactions. Finally, we conclude in section \ref{sec8}.

\section{The general construction\label{sec2}}

We begin by presenting the general case of our construction.  We will work in arbitrary dimension to start, and later specialize to the four dimensional case of interest.  The theory is that of a $(d-1)$-brane, with worldvolume coordinates $x^\mu$, moving in a  $D$-dimensional background, $D\geq d$, with coordinates $X^A$ and a fixed background metric $G_{AB}(X)$.  The dynamical variables include the brane embedding functions $X^A(x)$, $D$ functions of the world-volume coordinates $x^\mu$.   

We may construct the induced metric $\bar g_{\mu\nu}(x)$ via
\bea \label{inducedmetricdef}
\bar g_{\mu\nu}(x)&=& {\partial X^A\over\partial x^\mu} {\partial X^B\over\partial x^\nu}G_{AB}(X(x)) \ .
\eea
In addition to the induced metric, there are other geometric quantities  associated with the embedding, such as the extrinsic curvatures and the twist connection (see Appendix A of~\cite{Hinterbichler:2010xn} for a complete description of these quantities).

We would like the action on the world-volume to be invariant under re-parametrizations of the brane $x^\mu\rightarrow x^\mu-\xi^\mu(x)$, under which the embedding functions are scalars,
\be
\label{gaugetransformations} 
\delta_g X^A=\xi^\mu\partial_\mu X^A \ .
\ee
The induced metric~\eqref{inducedmetricdef} transforms as a tensor under these gauge transformations,
\be \delta_g \bar g_{\mu\nu}={\cal L}_\xi \bar g_{\mu\nu}= 
\xi^\lambda\partial_\lambda \bar  g_{\mu\nu}+\partial_\mu\xi^\lambda\, \bar g_{\lambda\nu}+\partial_\nu\xi^\lambda\, \bar g_{\mu \lambda} \ .\ee
Gauge invariance requires that the action be written as a diffeomorphism scalar, $F$, of the induced metric $\bar g_{\mu\nu}$, its covariant derivatives $\bar\nabla_\mu$, its curvature $\bar R^\rho_{\ \sigma\mu\nu}$, and the other induced quantities such as intrinsic curvature and twist (which we denote with ellipses),
\be
\label{generalaction} 
S= \int d^dx\ \sqrt{-\bar g}F\left(\bar g_{\mu\nu},\bar\nabla_\mu,\bar R^\rho_{\ \sigma\mu\nu},\cdots \right) \ .
\ee

In addition to the gauge symmetry of re-parametrization invariance, there can also be global symmetries.  If the bulk metric has a Killing vector $K^A(X)$, satisfying the Killing equation
\be \label{killingequation}
K^C\partial_C G_{AB}+\partial_AK^CG_{CB}+\partial_BK^CG_{AC}=0 \ ,
\ee
then the action will have a global symmetry under which the embedding scalars $X^A$ shift,
\be
\label{generalsym} 
\delta_K X^A=K^A(X) \ .
\ee
The induced metric~\eqref{inducedmetricdef} and other induced quantities, and therefore the general action~(\ref{generalaction}), are invariant under~(\ref{generalsym}).  

We may completely fix the re-parametrization freedom~\eqref{gaugetransformations} by fixing the unitary gauge
\be\label{gaugechoice}
X^\mu(x)=x^\mu,\ \ \ X^I(x)\equiv\pi^I(x) \ .
\ee
In this gauge, the world-volume coordinates of the brane are identified with the first $d$ of the bulk coordinates.  The remaining unfixed fields, $\pi^I(x)$, $I=1\cdots N$, where $N=D-d$ is the co-dimension of the brane, measure the transverse position of the brane.  The gauge fixed action is an action solely for $\pi$,
\be
\label{gaugefixedaction} 
S_{\bar g}= \int d^dx\ \left. \sqrt{-\bar g}F\left(\bar g_{\mu\nu},\bar\nabla_\mu,\bar R^{\alpha}_{\ \beta\mu\nu},\cdots\right)\right|_{X^\mu=x^\mu,\ X^I=\pi^I} \ .
\ee

The form of the global symmetries~(\ref{generalsym}) is altered once the gauge is fixed, because the gauge choice~\eqref{gaugechoice} is not generally preserved by the global symmetry.  The change induced by $K^A$ is $
\delta_ Kx^\mu=K^\mu(x,\pi),\ \delta_K\pi^I=K^I(x,\pi)$, so
to maintain the gauge~(\ref{gaugechoice}), we must simultaneously perform a compensating gauge transformation with the gauge parameter 
\be \label{compensating}
\xi_{\rm comp}^{\mu}=-K^\mu(x,\pi) \ .
\ee
The combined symmetry acting on the fields $\pi^I$ is now
\be
\label{gaugefixsym} 
(\delta_K+\delta_{{\rm comp}})\pi^I=-K^\mu(x,\pi)\partial_\mu\pi^I+K^I(x,\pi) \ ,
\ee
and is a global symmetry of the gauge fixed action~(\ref{gaugefixedaction}).

In addition to the induced metric~\eqref{inducedmetricdef}, we now introduce an additional world-volume metric $g_{\mu\nu}(x)$ onto the brane.  We demand that this obey the same transformation laws as $\bar g_{\mu\nu}$, and so we declare that it is invariant under the global symmetries~(\ref{generalsym}), but transforms as a tensor under~(\ref{gaugetransformations}),
\bea&& \delta_K g_{\mu\nu}=0, \\
&& \delta_g g_{\mu\nu}={\cal L}_\xi g_{\mu\nu}=\xi^\lambda\partial_\lambda g_{\mu\nu}+\partial_\mu\xi^\lambda\, g_{\lambda\nu}+\partial_\nu\xi^\lambda\, g_{\mu \lambda} \ .
\eea
We are now free to add to the action terms which are diffeomorphism scalars constructed from the intrinsic metric $g_{\mu\nu}$ and its associated covariant derivative and curvature,
\be
\label{gaction} 
S_{ g}= \int d^dx\  \sqrt{- g}F\left( g_{\mu\nu},\nabla_\mu, R^{\alpha}_{\ \beta\mu\nu}\right) \ ,
\ee
as well as terms that mix the intrinsic metric with the induced metric and other quantities\footnote{We have chosen $\sqrt{-g}$ to be the universal measure factor for each term in this part of the action.  This entails no loss of generality, since this choice may be traded for $\sqrt{-\bar g}$, or even a geometric mean such as $(-g)^{1/4}(-\bar g)^{1/4}$, by pulling out or absorbing powers of factors such as $\det\(g^{-1}\bar g\)$ into $F$. However,
whatever choice is made the absence of the ghost should be investigated, as we will do in subsequent sections.},
\be
\label{mixedaction} 
S_{\rm mix}= \int d^dx\  \sqrt{- g}F\left(g_{\mu\nu},\nabla_\mu, R^{\alpha}_{\ \beta\mu\nu},\bar g_{\mu\nu},\bar\nabla_\mu,\bar R^{\alpha}_{\ \beta\mu\nu},\cdots\right) \ .
\ee

Once the unitary gauge~\eqref{gaugechoice} is fixed, the fundamental fields of the theory are $g_{\mu\nu}$ and $\pi^I$, and the global symmetries act as
\bea 
\label{unitmetricsym} &&\delta g_{\mu\nu}=-K^\lambda(x,\pi)\partial_\lambda g_{\mu\nu}-\partial_\mu \left[ K^\lambda(x,\pi)\right]\, g_{\lambda\nu}-\partial_\nu \left[K^\lambda(x,\pi)\right]\, g_{\mu \lambda}\,, \\ \label{unitpisym}
&&\delta \pi^I=-K^\mu(x,\pi)\partial_\mu\pi^I+K^I(x,\pi) \ .
\eea
(Here, we must act by the derivatives on the argument of $\pi(x)$ within $K^\mu$, i.e. $\partial_\mu \left[ K^\lambda(x,\pi)\right]=\partial_\mu  K^\lambda+{\partial K^\lambda\over \partial\pi^I}\partial_\mu\pi^I$.) Note that in the unitary gauge the intrinsic metric $g_{\mu\nu}$ transforms non-trivially under the global symmetry, due to the compensating gauge transformation~\eqref{compensating}.  Under this (in general non-linear) transformation, the induced metric~\eqref{inducedmetricdef} transforms in precisely the same way as the intrinsic metric~\eqref{unitmetricsym},
\be 
\delta \bar g_{\mu\nu}=-K^\lambda(x,\pi)\partial_\lambda \bar g_{\mu\nu}-\partial_\mu \left[ K^\lambda(x,\pi)\right]\, \bar g_{\lambda\nu}-\partial_\nu \left[K^\lambda(x,\pi)\right]\, \bar g_{\mu \lambda} \ .
\ee 

This construction allows us to have scalars with galileon-like shift symmetries given by~\eqref{unitpisym}, coupled to a dynamical metric, which now carries a non-trivial transformation~\eqref{unitmetricsym} under the galilean symmetries.  It remains to ensure the final desired property --- that the action is free of ghosts.

\section{Ghost-free actions\label{sec3}}

If, as for the original galileons, the actions are to be free from extra Boulware-Deser-like~\cite{Boulware:1973my} degrees of freedom, they must take a specific form.  For the part of the action $S_g$, depending only on the dynamical metric $g_{\mu\nu}$, we know that the only possibilities giving second order equations of motion are the Einstein-Hilbert term, the cosmological constant, and the higher Lovelock invariants~\cite{Lanczos:1938sf,Lovelock:1971yv} if the brane has dimension $d>4$,
\be 
\label{Lagrangian}
S_g={1\over 2\kappa^2}\int d^dx\ \sqrt{-g}\left[ -2\Lambda+R[g]+\cdots \right] \ .
\ee

For the term $S_{\bar g}$, depending only on the induced metric $\bar g_{\mu\nu}$, the possibilities are the Lovelock terms of $\bar g$, as well as the boundary terms associated with Lovelock terms in the bulk, as detailed in~\cite{Hinterbichler:2010xn}.  For co-dimension 1, these are the Myers boundary terms~\cite{Myers:1987yn}.  In the unitary gauge~\eqref{gaugechoice}, these lead to the galileon terms for the single $\pi$ field~\cite{deRham:2010eu}.  For higher co-dimension, the surface terms are more limited and difficult to catalogue~\cite{Charmousis:2005ey,Charmousis:2005ez}.  In the unitary gauge they lead to the multi-field galileons~\cite{Hinterbichler:2010xn}.
In all cases, the leading term is the DBI term for the induced metric, which contains the kinetic term for the $\pi^I$ fields, and so we write this part of the action as 
\be 
S_{\bar g}=-T \int d^dx\ \sqrt{-\bar g}+\cdots \ ,
\ee
where $T$ is a constant of mass dimension $d$, and the ellipses denote the possible higher order Lovelock and boundary terms.

For the mixed terms, it is not immediately obvious what the most general ghost-free terms are.  However, if we restrict to terms depending only on $g_{\mu\nu}$ and $\bar g_{\mu\nu}$, with no higher derivatives, we can take a clue from the dRGT theory~\cite{deRham:2010ik,deRham:2010kj} of massive gravity and the related models of bi-gravity~\cite{Hassan:2011zd}, all of which have been shown to be ghost free~\cite{Hassan:2011hr,Hassan:2011tf,Hassan:2011ea,deRham:2011qq,deRham:2011rn,Mirbabayi:2011aa,Hinterbichler:2012cn}.  These models contain interaction terms between two metrics --- the second metric is a fixed fiducial metric in the case of massive gravity and a dynamical second metric in the case of bi-gravity.  In this paper we will choose the form of the interactions to be the same, but the second metric will be the induced metric \eqref{inducedmetricdef}, containing the $\pi^I$ degrees of freedom.  
The interactions can be constructed through the tensor  ${\cal K}^\mu_{\ \nu}=\delta^\mu_{\ \nu}-\sqrt{g^{\mu\lambda}\bar g_{\lambda\nu}}$, in terms of which the relevant piece of the action is given as follows 
\be 
\label{krep}
S_{mix} =-{M_{\rm Pl}^2\over 2}\int d^4 x\sqrt{-g}~{{m^2}\over 4}\sum_{n=2}^4\alpha_n S_n\(\cal K\)\,,
\ee
where $S_n(M)$, $0\leq n\leq d$, for a $d\times d$ matrix $M^\mu_{\ \nu}$, are the elementary symmetric polynomials\footnote{Our anti-symmetrization weight is $[\mu_1\ldots\mu_n]={1\over n!}(\mu_1\cdots\mu_n+\cdots)$.  The appearance of the symmetric polynomials and their relation to the absence of ghosts can be naturally seen in the vielbein formulation of the theory~\cite{Hinterbichler:2012cn}.  See appendix A of ~\cite{Hinterbichler:2012cn} for more details on the symmetric polynomials.} 
\be
\label{e}
S_n(M) = 
M^{[ \mu_1}_{\ \mu_1}\cdots M^{ \mu_n]}_{\ \mu_n}\,,
\ee
and $\sqrt{g^{-1}\bar g}$ is the matrix square root of the matrix $g^{\mu\sigma}\bar g_{\sigma \nu}$.  The $\alpha_{3,4}$ are free coefficients, while $\alpha_2=-8$ is required for the correct normalization of the graviton mass. It is often convenient to work in terms of the expanded action
\be 
S_{\rm mix}=-{M_{\rm Pl}^2\over 2} \int d^dx  \sqrt{-g}\,{{m^2}\over 4}\sum_{n= 0}^{d}\beta_nS_n(\sqrt{g^{-1}\bar g}) \ ,
\label{sintterm}\ee
where $\beta_n$ can be expressed in terms of the two free parameters $\alpha_{3,4}$. We will use the latter representation of the mixing terms below. Note that the $n=0$ and $n=d$ terms in the latter sum are redundant, since these reproduce the cosmological constant $\sqrt{-g}$ and DBI term $\sqrt{-\bar g}$ respectively.

\section{Co-dimension zero: massive gravity\label{sec4}}

Our construction contains ghost-free dRGT massive gravity as a special case.  When the co-dimension is zero, we are embedding a $d$-dimensional world-volume into 
a bulk space of the same dimension.  The fixed bulk metric $G_{\mu\nu}$ therefore has the same dimension as the brane metric.  

In unitary gauge there are no $\pi$ fields, and the induced metric is the bulk metric,
\be 
\bar g_{\mu\nu}(x)=G_{\mu\nu}(x) \ ,
\ee
so that the global symmetries are the Killing vectors $\xi^\mu(x)$ of $G_{\mu\nu}(x)$: $\xi^\lambda\partial_\lambda G_{\mu\nu}+\partial_\mu\xi^\lambda\, G_{\lambda\nu}+\partial_\nu\xi^\lambda\, G_{\mu \lambda}=0.$  The intrinsic metric then transforms linearly as a tensor,
\be 
\label{globalmassive} \delta g_{\mu\nu}=\xi^\lambda\partial_\lambda g_{\mu\nu}+\partial_\mu\xi^\lambda\, g_{\lambda\nu}+\partial_\nu\xi^\lambda\, g_{\mu \lambda} \ .
\ee

The action $S_{\bar g}$ contains no dynamical variables and can be dropped.  If the bulk metric is flat, $G_{\mu\nu}=\eta_{\mu\nu}$, the action $S_{g}+S_{\rm mix}$ is precisely the Lorentz invariant dRGT massive gravity of~\cite{deRham:2010ik,deRham:2010kj}, in the unitary gauge.  Further, the Poincare invariance of these actions comes from~(\ref{globalmassive}).  For a general bulk metric, the 
theory is that of massive gravity with a general reference metric~\cite{Hassan:2011tf}, and the global symmetries~(\ref{globalmassive}) are precisely the isometries of the reference metric.  These theories are all ghost free~\cite{Hassan:2011hr,Hassan:2011tf,Hassan:2011ea}, meaning that they propagate, non-linearly, precisely the number of degrees of freedom of a massive graviton and no more.

Away from the unitary gauge, we have 
\bea 
\label{inducedmetricdef2}
\bar g_{\mu\nu}(x)&=& {\partial X^\rho\over\partial x^\mu} {\partial X^\sigma\over\partial x^\nu}G_{\rho\sigma}(X(x)) \ ,
\eea
which is nothing but the St\"uckelberg replacement used to restore diffeomorphism invariance to massive gravity~\cite{Siegel:1993sk,ArkaniHamed:2002sp}.

\section{Flat bulk case\label{sec5}}

We now return to general co-dimension $N$, but specialize to a flat bulk metric $G_{AB}=\eta_{AB}$.  The isometries are the Poincare transformations of the bulk,
\be 
\label{poincaretransformations} 
\delta_PX^A=K^A(X)=\omega^A_{\ B}X^B+\epsilon^A \ ,
\ee
where $\epsilon^A$ and the antisymmetric matrix $\omega^A_{\ B}$ are the infinitesimal parameters of the bulk translations and Lorentz transformations respectively.

The unitary gauge~\eqref{gaugechoice} is not in general preserved by the Poincare transformations, 
but the gauge is restored by making the compensating gauge transformation, $\delta_gX^\mu=\xi^\nu\partial_\nu x^\mu=\xi^\mu$, with the choice 
\be 
\xi^\mu_{\rm comp}=-\omega^\mu_{\ \nu}x^\nu-\omega^\mu_{\ I}\pi^I-\epsilon^\mu \ .
\ee
The combined transformation $\delta_{P'}=\delta_P+\delta_g$ then leaves the gauge fixing intact and is a symmetry of the gauge fixed action.  This symmetry acts on the remaining fields as
\be 
\label{multiinternalpoincare}
\delta_{P'}\pi^I=-\omega^\mu_{\ \nu}x^\nu\partial_\mu\pi^I-\epsilon^\mu\partial_\mu \pi^I+\omega^I_{\ \mu}x^\mu+\omega^{\ \mu}_{ J}\pi^J\partial_\mu\pi^I+\epsilon^I+\omega^I_{\ J}\pi^J \ ,
\ee
where the first two terms in this expression are unbroken spacetime rotations and translations, respectively.  The second two terms are a DBI symmetry corresponding to the broken 
boosts in the extra dimensional directions (which becomes the galileon symmetry for small $\pi^I$).  The fifth term is a shift symmetry, corresponding to  
broken translations into the transverse directions.   Finally, the last term is the unbroken $SO(N)$ symmetry in the transverse directions, which appears as an internal rotation among the $\pi$ fields.  The symmetry breaking pattern is $ISO(1,D-1)\rightarrow ISO(1,d-1)\times SO(N)$.

The induced metric in unitary gauge is 
\be 
\label{inducedpoinc}
\bar g_{\mu\nu}=\eta_{\mu\nu}+\partial_\mu\pi^I \partial_\nu\pi_I \ ,
\ee
and using~\eqref{unitmetricsym}, we can determine how the global symmetries extend to the metric.  The metric transforms linearly, as a tensor, under the unbroken $d$-dimensional Poincare symmetry, and is invariant under the unbroken $SO(N)$ internal symmetry.  The broken DBI shift symmetries, on the other hand, extend non-trivially to the dynamical metric,
\be  
\delta g_{\mu\nu}=\omega_{I}^{\ \lambda}\pi^I\, \partial_\lambda g_{\mu\nu}+\omega_I^{\ \lambda}\partial_\mu\pi^I\, g_{\lambda\nu}+\omega_I^{\ \lambda}\partial_\nu\pi^I\, g_{\mu \lambda} \ .
\ee
This is a diffeomorphism with parameter $\xi^\mu=\omega_{I}^{\ \mu}\pi^I$, since its origin was nothing but a compensating diffeomorphism to restore unitary gauge.  The induced metric~\eqref{inducedpoinc}  transforms in the same way, as 
\be  
\delta \bar g_{\mu\nu}=\omega_{I}^{\ \lambda}\pi^I\, \partial_\lambda \bar g_{\mu\nu}+\omega_I^{\ \lambda}\partial_\mu\pi^I\, \bar g_{\lambda\nu}+\omega_I^{\ \lambda}\partial_\nu\pi^I\, \bar g_{\mu \lambda} \ .
\ee

In summary, in the unitary gauge the galileons $\pi^I$ and the intrinsic metric $g_{\mu\nu}$ transform as tensors under the unbroken $d$ dimensional Poincare symmetry, as a vector and singlet respectively under $SO(N)$, and as follows under the broken galileon symmetries,
\bea 
&&\delta \pi^I=\omega^I_{\ \mu}x^\mu+\omega^{\ \mu}_{ J}\pi^J\partial_\mu\pi^I+\epsilon^I \ , \nn \\
&& \delta g_{\mu\nu}=\omega_{I}^{\ \lambda}\pi^I\, \partial_\lambda g_{\mu\nu}+\omega_I^{\ \lambda}\partial_\mu\pi^I\, g_{\lambda\nu}+\omega_I^{\ \lambda}\partial_\nu\pi^I\, g_{\mu \lambda} \ . 
\label{transsumm}
\eea

This is how the DBI galileon symmetry extends to the metric.  It is a global symmetry.  In the unitary gauge we are working in, there is no diffeomorphism invariance (assuming there are interaction terms $S_{\rm mix}$ in the action).  As a consequence, the graviton described by $g_{\mu\nu}$ will be massive.  In this sense, it is natural for the galileons to couple to a massive graviton.

If we choose not to go to unitary gauge, diffeomorphism invariance on the brane then remains intact.  In this case we have the St\"uckelberg fields $X^\mu$, and the induced metric takes the form 
\be 
\label{inducedstukel}
\bar g_{\mu\nu}={\partial X^\rho\over\partial x^\mu} {\partial X^\sigma\over\partial x^\nu}\eta_{\rho\sigma}+\partial_\mu\pi^I \partial_\nu\pi_I \ .
\ee
Because we now still have diffeomorphism invariance, the induced and intrinsic metrics are invariant under the global symmetries~\eqref{poincaretransformations}, and transform as tensors under diffeomorphisms.  The fields $X^\mu$ and $\pi^I$ are scalars under the diffeomorphisms, and transform together as~\eqref{poincaretransformations} under the global symmetries.

\section{Small field expansions and decoupling limits\label{sec6}}

If the action has a background solution $g_{\mu\nu}=\eta_{\mu\nu}$ for the intrinsic metric, we may expand about it in fluctuations $h_{\mu\nu}$,
\be 
g_{\mu\nu}=\eta_{\mu\nu}+h_{\mu\nu} \ .
\ee
The unitary gauge non-linear transformation laws~\eqref{transsumm} expanded around this background are then
\bea 
\delta \pi^I&=&\omega^I_{\ \mu}x^\mu+\omega^{\ \mu}_{ J}\pi^J\partial_\mu\pi^I+\epsilon^I,   \nn \\
 \delta h_{\mu\nu}&=&\omega_{I \mu}\partial_\nu\pi^I+\omega_{I \nu}\partial_\mu\pi^I+\omega_{I}^{\ \lambda}\pi^I\, \partial_\lambda h_{\mu\nu}+\omega_I^{\ \lambda}\partial_\mu\pi^I\, h_{\lambda\nu}+\omega_I^{\ \lambda}\partial_\nu\pi^I\, h_{\mu \lambda} \ ,
\label{nonlintrans}
\eea
and we see that the metric fluctuations must transform along with the galileon fields.

Defining the fluctuation around the induced metric via
\be 
\label{fluctdef} H_{\mu\nu}=g_{\mu\nu}-\bar g_{\mu\nu} \ ,
\ee
we have in unitary gauge
\be 
H_{\mu\nu}=h_{\mu\nu}-\partial_\mu \pi^I\partial_\nu \pi_I \ .
\ee 

To perform the St\"uckelberg expansion, we simply leave the gauge unfixed, so that the induced metric takes the form~\eqref{inducedstukel}.  The fluctuation~(\ref{fluctdef}) can then be written as
\be 
H_{\mu\nu}=h_{\mu\nu}+\eta_{\mu\nu}-\partial_\mu X^\rho\partial_\nu X^\sigma\eta_{\rho\sigma}-\partial_\mu \pi^I\partial_\nu \pi_I \ .
\ee
As in massive gravity, we can then introduce another St\"uckelberg field $\phi$ to deal with the longitudinal mode through the following replacement, by expanding $X^\mu$ around its unitary gauge value,
\be 
X^\rho=x^\rho+A^\rho-\eta^{\rho\sigma}\partial_\sigma\phi \ .
\ee
The action then has the infinitesimal gauge transformations,
\bea
\delta h_{\mu\nu}&=&\partial_\mu \xi_\nu+\partial_\nu \xi_\mu+{\cal L}_\xi h_{\mu\nu} \ , \nn\\
 \delta A_\mu&=&\partial_\mu\Lambda-\xi_\mu+\xi^\nu\partial_\nu A_\mu \ , \nn\\ 
 \delta \phi&=& \Lambda \ ,\nn \\
 \delta \pi^I&=&\xi^\mu\partial_\mu\pi^I \ . 
 \label{gaugesyms3}
\eea
As is usually done in massive gravity, we ignore the vector mode $A^\mu$ which carries 
the helicity one components of the massive graviton at high energy (a consistent truncation), 
since these do not generally couple to matter at the linearized level. Putting them to zero 
is consistent with the equations of motion. Moreover, there is an enhanced $U(1)$ symmetry for this 
vector in the decoupling limit that guaranties that it propagates two degrees of freedom.  
We then have the replacement
\be
\label{flucfullreplace} 
H_{\mu\nu}=h_{\mu\nu}+2\partial_\mu\partial_\nu\phi-\partial_\mu\partial_\lambda\phi\, \partial_\nu\partial^\lambda\phi-\partial_\mu \pi^I\partial_\nu \pi_I \ .
\ee

For generic choices of the action, these theories describe a massive graviton coupled to the galileon fields $\pi^I$, with coupling such that the unitary gauge action is invariant under the galileon symmetries~\eqref{transsumm}.  Away from the unitary gauge on the other hand, the longitudinal mode of the massive graviton is described by the scalar $\phi$, which appears in addition to the $\pi^I$ when we restore the diffeomorphism invariance by not fixing unitary gauge.

For definiteness, we now focus on $d=4$.   The action is
\be 
\label{fullaction}
S_g+S_{\rm mix}+S_{\bar g}=\int d^4x \Bigg \{ \ {M_{\rm Pl}^2\over 2}\sqrt{-g}\left[R[g]-{{m^2}\over 4}\sum_{n=0}^4\beta_nS_n\(\sqrt{g^{-1}\bar g}\)\right]- M_{\rm Pl}^2m^2 \sqrt{-\bar g}\left(\lambda_0+\cdots\right)\Bigg\}  \ .
\ee
Here $\lambda_0$ and $\beta_n$ are order one dimensionless constants, independent of the mass scales $m$ and $M_{\rm Pl}$.  (We have chosen the mass scalings of the various terms so that there will be an interesting decoupling limit, with new ingredients beyond those appearing in the corresponding limit of massive gravity.)  The ellipses in the final term denote the possible higher-order Lovelock and boundary terms, each of which has its own independent order one coefficient and is suppressed by the mass scale $m$ (which will ensure that non-trivial galileon interactions survive the decoupling limit).

The coefficient $\beta_4$ is redundant with $\lambda_0$, and so we set $\beta_4=0$, and expand around flat space in the unitary gauge $g_{\mu\nu}=\eta_{\mu\nu}+h_{\mu\nu}$, $\bar g_{\mu\nu}=\eta_{\mu\nu}+\partial_\mu \pi^I\partial_\nu \pi_I$.  Tadpole cancellation, ensuring that flat space is a solution, requires 
\be
\label{condition1}
\beta_0 + 3 \beta_1 + 3 \beta_2 + \beta_3=0.
\ee   
At quadratic order, we find a Fierz-Pauli massive graviton.  One of the $\beta$'s can be absorbed into $m^2$, which we do by demanding 
\be
\label{condition2}
8 - \beta_1 - 2 \beta_2 - \beta_3=0.
\ee  
This ensures that the Fierz-Pauli massive graviton with a mass $m$ propagates at quadratic order.

In addition we find a kinetic term for the $\pi$ fields,
\be 
-\half M_{\rm Pl}^2m^2\left[\lambda_0+{1\over 8}\(\beta_1+3\beta_2+3\beta_3\)\right]\left(\partial\pi^I\right)^2 \ .
\ee
Thus, provided $\lambda_0+{1\over 8}\(\beta_1+3\beta_2+3\beta_3\)>0$, the theory propagates, in addition to the massive graviton,  $N$ healthy (having the correct sign kinetic term) scalars on this background.   

In the end, we have four free parameters for the flat space theory (plus those corresponding to the higher Lovelock terms): the graviton mass, two independent $\beta$'s corresponding to the two free parameters in the interactions of dRGT massive gravity, and a remaining independent parameter which corresponds to the strength of the $\pi$ kinetic term.  Note that, if we had been looking for curved (A)dS solutions, there would have been an additional parameter corresponding to the curvature of the background.

We now make the replacement~\eqref{flucfullreplace} and expand in powers of the fields.  After canonical normalization of the various kinetic terms via
\be 
\label{canonicalfields}\hat{h}\sim M_{\rm Pl}h\,,\ \ \ \hat \phi\sim m^2 M_{\rm Pl}\phi\,, \ \ \ \hat \pi \sim mM_{\rm Pl} \pi \ , 
\ee
one can examine the interaction terms to determine their associated interaction scales.

First focus on those interaction terms arising from $S_{\rm mix}$ (or from the DBI term in $S_{\bar g}$).  By virtue of the St\"uckelberg replacement~\eqref{flucfullreplace}, $\phi$ always appears with two derivatives, $\pi$ appears with one derivative, and $h$ appears with no derivatives.   A generic term with $n_h$ powers of $h_{\mu\nu}$, $n_\pi$ powers of $\pi^I$ and $n_\phi$ powers of $\phi$, reads
\be 
\sim m^2M_{\rm Pl}^2 h^{n_h} (\partial \pi)^{n_\pi} (\partial^2\phi)^{n_\phi}\sim \Lambda_\lambda^{4-n_h-2n_\pi-3n_\phi}  \hat h^{n_h} (\partial\hat \pi)^{n_\pi}(\partial^2\hat \phi)^{n_\phi} \ ,
\ee
where the scale suppressing the term is written as
\be
\label{scalesformula} \Lambda_{\lambda}=\left(M_{\rm Pl}m^{\lambda-1}\right)^{1/\lambda},\ \ \ \lambda={3n_\phi+2n_\pi+n_h-4\over n_\phi+n_\pi+n_h-2} \ .
\ee
Since we always assume $m<M_{\rm Pl}$,  the larger $\lambda$, the smaller is this scale.  Note that $n_\pi$ must be even, by virtue of the way it enters in~\eqref{flucfullreplace}, and we have $n_\phi+n_A+n_h\geq 3$, since we are only considering interaction terms.  
The terms suppressed by the smallest scale are $\phi$ self-interaction terms, $n_\pi=n_h=0$, which are suppressed by scales $\geq\Lambda_5$ and $<\Lambda_3$.  These terms, however, all cancel up to a total derivative due to the special structure of the ghost-free massive gravity interactions~\cite{deRham:2010ik,deRham:2010kj}.

The scale $\Lambda_3= \(M_{\rm Pl}m^2\)^{1/3}$ becomes the lowest scale, and is carried by terms of schematic form ($n\geq 1$)
\be
\label{firstlambda3term} \sim {1\over \Lambda_3^{n-1}}{\hat h(\partial^2\hat\phi)^n} \ ,
\ee
and\footnote{Had we not neglected the vector mode $A^\mu$, we would have seen that terms of the form $\partial A\partial A (\partial\partial\pi)^n/\Lambda^{3n}$ also survive in the limit at hand.}
\be
\label{secondlambda3term}
\sim {1\over \Lambda_3^{n}} {(\partial \hat \pi)^2(\partial^2\hat\phi)^n} \ . 
\ee
(Note that here we have included the term mixing $h$ and $\phi$ for $n=1$, even though it is a kinetic mixing term and not an interaction term, because it is from this mixing that $\phi$ acquires its kinetic term and its canonical normalization.) All other terms carry scales higher than $\Lambda_3$. There are a finite number of terms of the first type \eqref{firstlambda3term}, and they take the same form as they do in massive gravity~\cite{deRham:2010ik,deRham:2010kj}. However, there are an infinite number of terms of the second type \eqref{secondlambda3term}. 

We now return to the possibility of higher Lovelock terms in $S_{\bar g}$.  In the unitary gauge, before any decoupling limit, these are the DBI galileons~\cite{deRham:2010eu,Goon:2010xh} for $N=1$, and their multi-field generalizations for higher $N$~\cite{Hinterbichler:2010xn}.  As we have mentioned, these terms are suppressed by the scale $m$.  For example, for $N=1$ the leading Lovelock term beyond the DBI kinetic term is the trace of the extrinsic curvature, which leads to a cubic DBI galileon in the unitary gauge action, 
\be 
S_{\bar g}= -M_{\rm Pl}^2m^2\int d^4 x~\( \lambda_0\sqrt{1+(\partial\pi)^2}-{\lambda_1\over m}{1\over 1+(\partial\pi)^2}\partial_\mu\partial_\nu\pi\partial^\mu\pi\partial^\nu\pi+\cdots\).\label{galexamp} 
\ee
Away from the unitary gauge on the other hand, restoring the St\"uckelberg field and canonically normalizing via $\hat \pi \sim mM_{\rm Pl} \pi$, the various galileon terms yield interactions of the form
\be 
\sim M_{\rm Pl}^2m^2\(\partial^2\hat\phi\over \Lambda_3^3\)^{n_\phi}\(\partial\hat\pi\over M_{\rm Pl}m\)^{n_\pi-n}\(\partial^2\hat\pi\over \Lambda_3^3\)^{n},\ \ \ n_\pi -1> n=0,1,\cdots, 4 \ .
\ee
The terms with multiple powers of the the factor $\partial\hat\pi\over M_{\rm Pl}m$ arise from expanding out the square roots and denominators of~\eqref{galexamp}.  The interactions suppressed by $\Lambda_3$ are those with $n_\pi=n+2$.  Of these, the ones with $n_\phi=0$ are precisely the terms which survive the limit which recovers the normal galileons from the DBI galileons~\cite{deRham:2010eu}.  Those with $n_\phi>0$, of which there are an infinite number, describe the coupling of the galileons with the longitudinal mode of the graviton.  We work out these couplings for the case of the cubic galileon in Appendix \ref{cubicappend}.  All other terms are suppressed by scales larger than $\Lambda_3$.

If we take the decoupling limit,
\be 
M_{\rm Pl}\rightarrow \infty\,,\ \ \  m\rightarrow 0\,,\ \ \ \ \Lambda_3=(M_pm^2)^{1/3} \ {\rm fixed} \ ,
\ee
then all  interactions with scales greater than $\Lambda_3$ are set to zero, and the DBI-galileon terms become the normal galileons.  The only part of the Einstein-Hilbert action that survives the decoupling limit is the quadratic part,
\be 
S_g^{\rm }\supset\int d^4x\  \frac{M_{\rm Pl}^2}{8} h_{\mu\nu}{\cal E}^{\mu\nu,\alpha\beta}  h_{\alpha\beta} \ ,
\ee
where the kinetic operator for the graviton is that of linearized Einstein gravity\footnote{Explicitly,
\begin{equation}\label{Eoper}
  {\cal E}^{\mu\nu}_{\ \ \alpha\beta}\equiv\left(\eta^{(\mu}_{\ \alpha}\eta^{\nu)}_{\ \beta}-\eta^{\mu\nu}\eta_{\alpha\beta}\right)\square-2\partial^{(\mu}\partial_{(\alpha}\eta^{\nu)}_{\ \beta)} 
+\partial^\mu\partial^\nu\eta_{\alpha\beta}+\partial_\alpha\partial_\beta\eta^{\mu\nu}.
\end{equation}}.
The gauge symmetries~\eqref{gaugesyms3} in the decoupling limit become the linearized versions of those considered above
\bea
\delta h_{\mu\nu}&=&\partial_\mu \xi_\nu+\partial_\nu \xi_\mu\,, \ \ \  \delta A_\mu=\partial_\mu\Lambda\,, \nn\\ 
 \delta \phi&=&0\,, \ \ \  \delta \pi^I=0 \ . 
 \label{gaugesymsdec}
\eea

As an explicit example, consider the choice of coefficients 
$\beta_0=-24,\ \beta_1=8,\ \beta_2=\beta_3=\lambda_0=0$ so that the DBI term is gone, 
and set to zero all the higher Lovelock terms within $S_{\bar g}$\footnote{This 
corresponds to the choice $c_3={1\over 6}$, $d_5=-{1\over 48}$ in the notations 
of~\cite{deRham:2010ik,Hinterbichler:2011tt}.}.  This choice corresponds in pure 
massive gravity to the model which contains no nonlinear 
scalar-tensor interactions in the decoupling limit~\cite{deRham:2010ik}.  The corresponding 
dRGT action, written in terms of the quadratic, cubic, and quartic terms 
in the matrix ${\cal K}^\mu_\nu = \delta^\mu_{\ \nu}-\sqrt{g^{\mu\lambda}\bar g_{\lambda\nu}}$ and 
its traces \cite {deRham:2010kj}, can be rewritten for these particular coefficients 
in the form given in \cite {Hassan:2011vm} 
\be
\label{mm} 
S=\int d^4 x\ {M_{\rm Pl}^2\over 2}\sqrt{-g}\left[R[g]+{2{m^2}}\(3-{\rm Tr}\sqrt{g^{-1}\bar g}\)\right] \ ,
\ee
which is often referred as the ``minimal model''.  Unlike in pure massive gravity
though, here ${\bar g}$ is going to be an induced metric on the brane.
Thus the interactions we find will be entirely due to the extension of the theory we have developed here.

The decoupling limit Lagrangian is\footnote{For deriving the decoupling limit, it is convenient to write the Lagrangian in terms of the tensor  ${\cal K}^\mu_{\ \nu}=\delta^\mu_{\ \nu}-\sqrt{g^{\mu\lambda}\bar g_{\lambda\nu}}$,
${\cal L}={M_{\rm Pl}^2\over 2}\sqrt{-g}\left(R[g]+2m^2\(-1+ {\cal K}^\mu_{\ \mu}\)\right).$  Putting in the St\"uckelberg fields,
\be {\cal K}^\mu_{\ \nu}=\delta^\mu_{\ \nu}-\sqrt{\delta^\mu_{\ \nu}-g^{\mu\lambda}H_{\lambda \nu}},\ \ \ H_{\mu\nu}=h_{\mu\nu}+2\Pi_{\mu\nu}-\Pi^2_{\mu\nu}-P_{\mu\nu},\ee
with $\Pi_{\mu\nu}=\partial_\mu\partial_\nu\phi$, and $P_{\mu\nu}=\partial_\mu \pi^I\partial_\nu \pi_I$.  We then use the relation (in matrix notation with $\la\ \ra$ the trace) $\la \delta {\cal K}\ra =-\half\la (1-\Pi)^{-1}P\ra$ to find the terms involving $\pi$, and ${\cal K}=\Pi+\half h-{1\over 4} h\Pi-{1\over 4} \Pi h$ for the terms involving $h$.}
\be 
{\cal L}_{\rm dec}=\frac{M_{\rm Pl}^2}{8} h_{\mu\nu}{\cal E}^{\mu\nu,\alpha\beta}  h_{\alpha\beta} +{M_{\rm Pl}^2m^2}\left[\half h_{\mu\nu}\( \eta^{\mu\nu} \square\phi-\partial^{\mu}\partial^{\nu}\phi\)- \half {1\over\delta^\mu_\nu- \partial^\mu\partial_\nu\phi} \partial_\mu \pi^I\partial^\nu \pi_I\right] \ .
\ee
Diagonalizing via $h_{\mu\nu}=h'_{\mu\nu}+m^2\phi\, \eta_{\mu\nu}$, we find a free decoupled graviton, and coupled interacting scalars, 
\be 
{\cal L}_{\rm dec}=\frac{M_{\rm Pl}^2}{8} h'_{\mu\nu}{\cal E}^{\mu\nu,\alpha\beta}  h'_{\alpha\beta} +{M_{\rm Pl}^2m^2}\left[-{3\over 4}m^2 \(\partial\phi\)^2- \half{1\over\delta^\mu_\nu- \partial^\mu\partial_\nu\phi} \partial_\mu \pi^I\partial^\nu \pi_I\right] \ .
\label{lambda3decouplingfull2}
\ee
The terms involving $\pi^I$ are new to this model, and do not appear in pure massive gravity.  Note that, unlike massive gravity, there are an infinite number of scalar interaction terms that survive the decoupling limit.

\section{Equations of motion and ghosts\label{sec7}}

The equations of motion obtained from~\eqref{lambda3decouplingfull2} are not second order.  To see this, we need only expand to cubic order in the fields, ${\cal L}_{\rm cubic}\sim \partial_\mu\partial_\nu\phi \,\partial^\mu\pi^I\partial^\nu\pi_I$.  The $\phi$ equation of motion, for example, is third order, $\sim \partial_\mu\partial_\nu\( \partial^\mu\pi^I\partial^\nu\pi_I\)$.  

Higher order equations are generally associated with extra ghostly degrees of freedom.  In dRGT massive gravity, the decoupling limit is second order and contains no extra degrees of freedom, as it must since the entire theory has no such extra degrees of freedom.  The higher order equations we are finding here are naively worrisome, because if the decoupling limit contains extra degrees of freedom, the entire model is not ghost free.

However, higher order equations do not necessarily imply the existence of extra degrees of freedom.  As we will now show, the Lagrangian~\eqref{lambda3decouplingfull2} in fact contains no additional ghostly degrees of freedom, despite the higher derivatives.

\subsection{A toy example}

As a warmup to proving this, consider first the simpler $0+1$ dimensional version of the scalar part of the action~\eqref{lambda3decouplingfull2},
\be 
S=\int dt\bigg( \half \dot\phi^2 +\half {\dot\pi^2\over 1+\ddot\phi} \bigg)\ \ \ .
\label{toyaction}
\ee
This is a higher-derivative Lagrangian, and the equations of motion are naively fourth order (third order if we expand to cubic order in the fields, as we did in the full case above), 
\bea 
&&{\delta S\over \delta \pi}={d\over dt}\left[ {\dot\pi \over 1+\ddot\phi}\right]\,, \\
&&{\delta S\over \delta \phi}=\ddot \phi+\half{d^2\over dt^2}\left[  {\dot\pi^2\over (1+\ddot\phi)^2}\right] \ .
\eea
As with the full theory, this seems worrisome, since it raises the possibility of extra degrees of freedom which are ghosts.  

However, the number of initial data needed to solve this system is only four, consistent with there being only two degrees of freedom.  To see this, note that the $\pi$ equation implies ${\dot\pi \over 1+\ddot\phi}$ is a constant, which when substituted into the $\phi$ equation implies $\ddot\phi=0$.  Substituting this back into the $\pi$ equation then yields $\ddot\pi=0$.  Thus the equations above are, in fact, equivalent to the free field equations $\ddot\phi=\ddot\pi=0$, and there are therefore no extra degrees of freedom.  

Note that Ostragradskii's theorem~\cite{Ostrogradski} does not apply to~\eqref{toyaction} (or to the full model~\eqref{lambda3decouplingfull2}), since one of the conditions of the theorem, that the Lagrangian be 
non-degenerate in the higher derivatives ({\it i.e.}, the matrix obtained by variation of the action with respect to 
second derivatives be nondegenerate), is not satisfied.  

The absence of extra degrees of freedom can also be seen directly at the level of the action.  Starting with~\eqref{toyaction}, we introduce an auxiliary field $\sigma$ to render the action polynomial in the fields,
\be 
S=\int dt\left[ \half \dot\phi^2 -\half \sigma^2(1+\ddot\phi)+\dot\pi \sigma \right] .
\label{toyaction2}
\ee
The equation of motion for $\sigma$ is then $\sigma={\dot\pi\over 1+\ddot\phi}$, which when substituted back into~\eqref{toyaction2} recovers~\eqref{toyaction}.  Integrating by parts to remove the second derivatives from $\phi$, we find an equivalent first order action,
\be 
S=\int dt\left[ \half \dot\phi^2 -\half \sigma^2+\sigma\dot\sigma\dot\phi-\pi \dot\sigma \right] .
\label{toyaction3}
\ee

We now Legendre transform to find an equivalent Hamiltonian action.  The canonical momenta are
\be 
p_\phi=\dot\phi+\sigma\dot\sigma,\ \ \ p_\pi=0,\ \ \ p_\sigma=\sigma\dot\phi-\pi \ .
\ee
There is a primary constraint
\be
\label{constraint} p_\pi=0 \ ,
\ee
for which we introduce the multiplier $\lambda$, and the action then takes the form
\be 
S=\int dt\left[ p_\phi\dot\phi+p_\pi\dot\pi+p_\sigma\dot\sigma-\left(-{1\over 2\sigma^2}\left(p_\sigma+\pi\right)^2+{p_\phi\over \sigma}\left(p_\sigma+\pi\right)+{\sigma^2\over 2}\right)-\lambda p_\pi \right] .
\ee
It is now straightforward to see that the primary constraint~\eqref{constraint} generates a secondary constraint, leaving four phase space degrees of freedom, or two Lagrangian degrees of freedom.  

Alternatively, we may solve the primary constraint directly in the action, 
\be 
S=\int dt\left[ p_\phi\dot\phi+ p_\sigma\dot\sigma-\left(-{1\over 2\sigma^2}\left(p_\sigma+\pi\right)^2+{p_\phi\over \sigma}\left(p_\sigma+\pi\right)+{\sigma^2\over 2}\right) \right] .
\ee
Now $\pi$ is an auxiliary field and can be eliminated through its equation of motion, $\pi=p_\phi\sigma- p_\sigma$, leaving
\be 
S=\int dt\left[ p_\phi\dot\phi+ p_\sigma\dot\sigma-\left(\half p_\phi^2+\half\sigma^2\right) \right] .
\ee
Thus, we see explicitly that there are exactly two degrees of freedom, since renaming $ p_\sigma\rightarrow q$, $\sigma\rightarrow -p_q$, the action is equivalent to that of two free particles with positive energy,
\be 
S=\int dt\left[ p_\phi\dot\phi+p_q\dot q-\left(\half p_\phi^2+\half p_q^2\right) \right] .
\ee

\subsection{No extra degrees of freedom in the decoupling limit}

We now apply apply a Hamiltonian analysis to the full scalar action in the decoupling limit of the theory \eqref{lambda3decouplingfull2}.
Setting $3m^2/4\equiv 1/2$, and specializing to the case of a single $\pi$ field for simplicity, the action \eqref{lambda3decouplingfull2} we are studying is proportional to
\be\label{lagin}
S=\int d^4x\ \ \left[ -\half \(\partial\phi\)^2- \half  \partial_\mu \pi {1\over\delta^\mu_\nu- \partial^\mu\partial_\nu\phi}\partial^\nu \pi\right].
\ee

To eliminate the inverse powers of derivatives and work with a local action, we introduce an auxiliary vector field $\Omega^{\mu}$, and write the following equivalent action,
\be
\label{auxaction}
S=\int d^4x\ \ \left[ -\frac{1}{2}(\partial\phi)^2 +\frac{1}{2}\Omega_{\mu}\(\delta^\mu_\nu- \partial^\mu\partial_\nu\phi\)\Omega^{\nu}-\Omega^{\mu}\partial_{\mu}\pi\right] \ .
\ee
Integrating out $\Omega_\mu$ through its equations of motion recovers \eqref{lagin}.
Now we make a $(3+1)$ decomposition of the Lorentz indices, and do some integrations by parts to remove all the double time derivatives from $\phi$, 
\bea
S=\int d^4x\ \left[ \frac{1}{2}{\dot \phi}^2-\frac{1}{2}(\vec \nabla\phi)^2-\frac{1}{2}(\Omega^0)^2+\frac{1}{2}(\Omega^i)^2+\Omega^{0}{\dot \Omega}^{0}{\dot \phi}\right ] \nonumber \\
+\int d^4x\ \left [
\partial_i(\Omega^0\Omega^i){\dot \phi}-\frac{1}{2}\Omega^i\Omega^j\partial_i\partial_j\phi+{\dot \Omega}^{0}\pi-\Omega^i\partial_i\pi \right]\ .
\eea

As in the toy model, we now have a Lagrangian which has at most first time derivatives, 
so we may pass to a Hamiltonian form of the action in standard fashion.
The conjugate momenta are
\bea
p_{\phi} &=& {\dot \phi}+\Omega^{0}{\dot \Omega}^{0}+\partial_i(\Omega^0\Omega^i) \ , \\
p_{\Omega^0} &=& \Omega^0 {\dot \phi} +\pi \ , \\
p_{\pi} &=& 0 \ ,\\
p_{\Omega^i} &=& 0 \ .
\eea
From this, we see that we have the primary constraints,
\be  p_{\pi} = 0\,,\ \ \ \ p_{\Omega^i} = 0\,.\label{constraintsp}\ee
The Hamiltonian density, on the constraint surface, is
\bea
{\cal H}&=& p_\phi\dot\phi+p_{\Omega^0}\dot \Omega^0+p_\pi\dot\pi+p_{\Omega^i}\dot \Omega^i-{\cal L} \nn\\
&=&\frac{p_{\phi}}{\Omega^0}(p_{\Omega^0}-\pi) -\frac{1}{2(\Omega^0)^2}(p_{\Omega^0}-\pi)^2 + \frac{1}{2}(\nabla\phi)^2+\frac{1}{2}(\Omega^0)^2-\frac{1}{2}(\Omega^i)^2 \nn \\
 &&- \frac{1}{\Omega^0}\partial_i(\Omega^0\Omega^i)(p_{\Omega^0}-\pi)+ \frac{1}{2}\Omega^i\Omega^j\partial_i\partial_j\phi+\Omega^i\partial_i\pi \ . \label{ham}
\eea
Thus,  solving the constraint inside the action gives the following equivalent Hamiltonian form of the action 
\be S=\int d^4x\ \ \left[ p_\phi\dot\phi+p_{\Omega^0}\dot \Omega^0-{\cal H}(p_\phi, \phi,p_{\Omega^0},\Omega^0,\pi,\Omega^i)\right].\ee
We can now see that the system describes precisely two fields' worth of degrees of freedom: both $\pi$ and $\Omega^i$ appear algebraically and can be integrated out, leaving a Hamiltonian action depending only on the phase space variables $p_\phi, \phi,p_{\Omega^0},\Omega^0$, for which there are standard unconstrained first order Hamiltonian equations.

Doing this explicitly, we first integrate out $\pi$ with its equation of motion
\be
\pi=p_{\Omega^0}-\Omega^0p_{\phi}+\Omega^0\Omega^i\partial_i\Omega^0 \ ,
\ee
which, substituted back into the Hamiltonian density, yields
\bea
{\cal H}&=&\frac{p_{\phi}^2}{2} - (\partial_i\Omega^i)p_{\Omega^0}-(\Omega^i\partial_i\Omega^0)p_{\phi}+\frac{1}{2}(\Omega^i\partial_i\Omega^0)^2+\frac{1}{2}(\nabla\phi)^2 \nn\\
&+& \frac{1}{2}(\Omega^0)^2-\frac{1}{2}(\Omega^i)^2+\frac{1}{2}\Omega^i\Omega^j\partial_i\partial_j\phi \ .
\eea
Next we eliminate $\Omega^i$ through its equation of motion
\be
\Omega^i=(A^{-1})^{ij}(\partial_jp_{\Omega^0}-p_{\phi}\partial_j\Omega^0) \ ,
\ee
where $A^i_j \equiv \delta^i_j-(\partial^i\Omega^0)\partial_j\Omega^0-\partial^i\partial_j\phi \ ,$ 
giving the Hamiltonian 
\be
{\cal H}=\frac{p_{\phi}^2}{2}+\frac{1}{2}(\partial_ip_{\Omega^0}-p_{\phi}\partial_i\Omega^0)(A^{-1})^{ij}(\partial_jp_{\Omega^0}-p_{\phi}\partial_j\Omega^0)+\frac{1}{2}(\nabla\phi)^2 + \frac{1}{2}(\Omega^0)^2 \ .
\ee
This describes two fields with nonlinear, spatially non-local interactions.

At the linear level, we have
\be
{\cal H}=\frac{p_{\phi}^2}{2}+\frac{1}{2}(\nabla p_{\Omega^0})^2+\frac{1}{2}(\nabla\phi)^2 + \frac{1}{2}(\Omega^0)^2 \ ,
\ee
so if we redefine $p_{\Omega^0}\equiv\chi$ and $\Omega^0\equiv -p_{\chi}$, we can see that this describes two ghost-free, non-tachyonic modes
\be
{\cal H}=\frac{p_{\phi}^2}{2}+\frac{p_{\chi}^2}{2}+\frac{1}{2}(\nabla\phi)^2 +\frac{1}{2}(\nabla\chi)^2 \ .
\ee

As already noted above, we have assumed that the vector mode also present in the decoupling limit does not get excited\footnote{Note that the presence of the infinite number of interactions of the form $\partial A\p A(\p\p \pi)^n$ in the decoupling limit is also essential for the (nonlinearly realized) invariance under the broken bulk Lorentz generators in \eqref{poincaretransformations}. While the inhomogeneous piece in the galileon transformation is $\delta\pi=\omega_\mu x^\mu$, the vector St\"uckelberg mode shifts under this generator as $\delta A_\mu=-\omega_\mu\pi$. The infinite number of these terms then should relate by this symmetry to the infinite number of scalar interactions found in \eqref{lambda3decouplingfull2}. }, so that we may set $A^\mu=0$. This is a consistent truncation of the action, since $A^\mu$ only enters at the quadratic level in the action. (It is also consistent quantum mechanically, in the decoupling limit, due to the $Z_2$ symmetry $A^\mu\to -A^\mu$.) For arbitrary excitations of this mode, our Hamiltonian treatment has to be modified. We have no handle on the infinite number of $\partial A\p A(\p\p \pi)^n$ interactions present in the decoupling limit (even in dRGT gravity their form is in general not known beyond the cubic order \cite{deRham:2010gu}, see however \cite{Mirbabayi:2011aa,Koyama:2011wx}). However, it is plausible 
to expect that the presence of the vector mode does not spoil the ghost-free property of the 
decoupling limit because of the enhanced $U(1)$ symmetry of the limiting action; 
this is precisely what happens in ghost-free massive gravity. 

If curvature invariants composed of the induced metric are added as implied by an ellipsis in \eqref{Lagrangian}, the proof of ghost freedom becomes increasingly difficult; the special structure of the resultant decoupling limit interactions however leads us to conjecture that the absence of the extra degrees of freedom carries through to this case as well. The simplest such interaction, corresponding to a cubic galileon for $\pi$ in the decoupling limit, is considered in Appendix \ref{cubicappend}.

\section{Summary and prospects\label{sec8}}

We have introduced a model which couples a scalar $\pi^I$ to a dynamical metric in a manner which respects the galileon symmetries.  The metric to which the galileons couple is a massive graviton.   The model can be considered as an extension to higher co-dimension of ghost-free dRGT massive gravity, or as an extension of the brane construction of the galileons where a dynamical metric is added to the brane.

We have derived explicitly the decoupling limit of the model around flat space, for a specific choice of parameters, and have shown that there are no ghosts.  We have not proven that there are no ghosts beyond the decoupling limit, though we expect that there should not be, since the model is based on the ghost-free constructions of dRGT massive gravity and galileon theories.

This model should provide a completely consistent framework within which to investigate the implications of galileon invariance in, for example, cosmology.

\vskip.5cm

\bigskip
{\bf Acknowledgements}: 
We would like to thank Claudia de Rham and Andrew Tolley for helpful discussions.
GG is supported by NSF grant PHY-0758032 and NASA grant NNX12AF86G S06, 
and in part, by the Government of Canada through Industry Canada, and by the Province 
of Ontario through the Ministry of Research and Information.
K.H. is supported by funds provided by the University of Pennsylvania, and M.T. is supported in part by the US Department of Energy, and NSF grant PHY-0930521. J.K. is supported by NASA ATP grant NNX11AI95G, the Alfred P. Sloan Foundation and NSF CAREER Award PHY-1145525.  D.P. is supported by the U.S. Department of Energy under contract No. DOE-FG03-97ER40546.

\appendix

\section{Adding a cubic galileon\label{cubicappend}}

In this appendix, specializing to the case of a single extra dimension for simplicity,  we derive the decoupling limit interactions resulting from adding the extrinsic curvature term $M_{\rm Pl}^2 m \int d^4 x\sqrt{-\bar g} ~K(\bar g)$ to the r.h.s. of \eqref{Lagrangian}.

As above, we we work in the gauge $X^\mu=X^\mu(x),~ X^5=\pi(x)$, so that the function $\Phi(x^A)\equiv \pi\(x(X)\)-X^5=0$ defines the embedding. The vector $n_A$, normal to the brane has the following components,
\be
n_A=\frac{\p_A \Phi}{|\eta^{AB}\p_A\Phi~\p_B\Phi|^{1/2}} \Rightarrow n_\mu=\frac{\bar\p_\mu\pi}{(1+\bar\p ^\alpha\pi\bar\p_\alpha\pi)^{1/2}}, \quad n_5 = -\frac{1}{(1+\bar\p ^\alpha\pi\bar\p_\alpha\pi)^{1/2}}\,,
\ee
where the operator $\bar \p$ denotes differentiation with respect to the bulk coordinate,
\be\bar \p_\mu = \frac{\p x^\rho}{•\p X^\mu } \frac{\p}{\p x^\rho} = \(\p_\rho X^\mu \)^{-1}\frac{\p}{\p x^\rho} \equiv A_\mu^{~\rho}(x)\p _\rho\,.
\ee
 The trace of the extrinsic curvature is then given by
 \be
 K^\mu_\mu=n_{A,B}~ e^A_\mu ~e^B_\nu ~\bar g^{\mn}=n_{A,B}~\eta^{AB} = \frac{1}{(1+\bar\p ^\alpha\pi\bar\p_\alpha\pi)^{1/2}}\(\bar \Box\pi-\frac{\bar\p_\mu\pi\bar\p_\nu\pi\bar\p^\mu\bar\p^\nu\pi}{1+\bar\p ^\alpha\pi\bar\p_\alpha\pi}\)\,.
 \ee
The last step is to evaluate $\sqrt{-\bar g}$,
\be
\bar g_{\mn}=\p_\mu X^\alpha\p_\nu X^\beta \eta_{\alpha\beta}+\p_\mu\pi\p_\nu\pi\,,
\ee
which, multiplied by two factors of the operator $A$ on both sides, gives
\be
\text{det}\(A_\lambda^{~\mu} A _\rho^{~\nu} \bar g_{\mn}\)=\text{det} \(\eta_{\lambda\rho}+\bar\p_\lambda\pi\bar\p_\rho\pi\)=-1-(\bar\p\pi)^2\Rightarrow \bar g=-\(1+(\bar\p\pi)^2\) \text{det} \(A^{-2}\) \nn,
\ee
so that we have\footnote{ Of course, noting that $d^4 x~ \text{det}(\p X)=d^4X$, this could be directly obtained by transforming the corresponding unitary gauge expression under a diffeomorphism $x^\mu\to X^\mu$.} ,
\be
\mpl^2 m \int d^4 x~\sqrt{-\bar g}~ K = \mpl^2 m \int d^4 x ~ \text{det}(\p X)~\(\bar \Box\pi-\frac{\bar\p_\mu\pi\bar\p_\nu\pi\bar\p^\mu\bar\p^\nu\pi}{1+\bar\p ^\alpha\pi\bar\p_\alpha\pi}\)\,.
\ee
In the weak field, and the decoupling limits, this leads to an extra cubic galileon with the 5D derivatives $\p\to\bar\p$ instead of ordinary ones in \eqref{lambda3decouplingfull2},
\be
\L_{\text{dec}} \supset -\mpl^2 m \int d^4 x ~\text{det}\(1-\partial\partial\phi\)~\bar\p_\mu\pi\bar\p_\nu\pi\bar\p^\mu\bar\p^\nu\pi\,,
\ee
where (in the decoupling limit) $\bar \p_\mu \equiv \big [ \(1-\partial\partial\phi\)^{-1}\cdot \p\big ]_\mu$~. In $(0+1)$ one dimension, the latter expression becomes a surface term, as can be seen by reparametrizing the time coordinate $t\to t'=\int dt (1-\ddot \phi)$.

\bibliographystyle{utphys}
\addcontentsline{toc}{section}{References}
\bibliography{dynamicalmetricgalileon9}

\providecommand{\href}[2]{#2}\begingroup\raggedright\begin{thebibliography}{10}

\bibitem{Luty:2003vm}
M.~A. Luty, M.~Porrati, and R.~Rattazzi, ``{Strong interactions and stability
  in the DGP model},'' {\em JHEP} {\bfseries 0309} (2003) 029,
\href{http://arxiv.org/abs/hep-th/0303116}{{\ttfamily arXiv:hep-th/0303116
  [hep-th]}}.

\bibitem{Nicolis:2008in}
A.~Nicolis, R.~Rattazzi, and E.~Trincherini, ``{The galileon as a local
  modification of gravity},''
\href{http://arxiv.org/abs/0811.2197}{{\ttfamily arXiv:0811.2197 [hep-th]}}.

\bibitem{Deffayet:2009mn}
C.~Deffayet, S.~Deser, and G.~Esposito-Farese, ``{Generalized Galileons: All
  scalar models whose curved background extensions maintain second-order field
  equations and stress-tensors},''
  \href{http://dx.doi.org/10.1103/PhysRevD.80.064015}{{\em Phys. Rev.}
  {\bfseries D80} (2009) 064015},
\href{http://arxiv.org/abs/0906.1967}{{\ttfamily arXiv:0906.1967 [gr-qc]}}.

\bibitem{Deffayet:2009wt}
C.~Deffayet, G.~Esposito-Farese, and A.~Vikman, ``{Covariant Galileon},''
  \href{http://dx.doi.org/10.1103/PhysRevD.79.084003}{{\em Phys. Rev.}
  {\bfseries D79} (2009) 084003},
\href{http://arxiv.org/abs/0901.1314}{{\ttfamily arXiv:0901.1314 [hep-th]}}.

\bibitem{deRham:2010eu}
C.~de~Rham and A.~J. Tolley, ``{DBI and the Galileon reunited},''
  \href{http://dx.doi.org/10.1088/1475-7516/2010/05/015}{{\em JCAP} {\bfseries
  1005} (2010) 015},
\href{http://arxiv.org/abs/1003.5917}{{\ttfamily arXiv:1003.5917 [hep-th]}}.

\bibitem{Goon:2012dy}
G.~Goon, K.~Hinterbichler, A.~Joyce, and M.~Trodden, ``{Galileons as
  Wess-Zumino Terms},'' {\em JHEP} {\bfseries 1206} (2012) 004,
\href{http://arxiv.org/abs/1203.3191}{{\ttfamily arXiv:1203.3191 [hep-th]}}.

\bibitem{Burrage:2011bt}
C.~Burrage, C.~de~Rham, and L.~Heisenberg, ``{de Sitter Galileon},''
  \href{http://dx.doi.org/10.1088/1475-7516/2011/05/025}{{\em JCAP} {\bfseries
  1105} (2011) 025},
\href{http://arxiv.org/abs/1104.0155}{{\ttfamily arXiv:1104.0155 [hep-th]}}.

\bibitem{Goon:2011qf}
G.~Goon, K.~Hinterbichler, and M.~Trodden, ``{Symmetries for Galileons and DBI
  scalars on curved space},''
\href{http://arxiv.org/abs/1103.5745}{{\ttfamily arXiv:1103.5745 [hep-th]}}.

\bibitem{Goon:2011uw}
G.~Goon, K.~Hinterbichler, and M.~Trodden, ``{General Embedded Brane Effective
  Field Theories},''
\href{http://arxiv.org/abs/1103.6029}{{\ttfamily arXiv:1103.6029 [hep-th]}}.

\bibitem{Goon:2011xf}
G.~Goon, K.~Hinterbichler, and M.~Trodden, ``{Galileons on Cosmological
  Backgrounds},'' \href{http://dx.doi.org/10.1088/1475-7516/2011/12/004}{{\em
  JCAP} {\bfseries 1112} (2011) 004},
\href{http://arxiv.org/abs/1109.3450}{{\ttfamily arXiv:1109.3450 [hep-th]}}.

\bibitem{Goon:2012mu}
G.~Goon, K.~Hinterbichler, A.~Joyce, and M.~Trodden, ``{Gauged Galileons From
  Branes},'' \href{http://arxiv.org/abs/1201.0015}{{\ttfamily arXiv:1201.0015
  [hep-th]}}.
5 pages.

\bibitem{deRham:2010ik}
C.~de~Rham and G.~Gabadadze, ``{Generalization of the Fierz-Pauli Action},''
  \href{http://dx.doi.org/10.1103/PhysRevD.82.044020}{{\em Phys. Rev.}
  {\bfseries D82} (2010) 044020},
\href{http://arxiv.org/abs/1007.0443}{{\ttfamily arXiv:1007.0443 [hep-th]}}.

\bibitem{deRham:2010kj}
C.~de~Rham, G.~Gabadadze, and A.~J. Tolley, ``{Resummation of Massive
  Gravity},'' \href{http://dx.doi.org/10.1103/PhysRevLett.106.231101}{{\em
  Phys.Rev.Lett.} {\bfseries 106} (2011) 231101},
\href{http://arxiv.org/abs/1011.1232}{{\ttfamily arXiv:1011.1232 [hep-th]}}.

\bibitem{Hinterbichler:2011tt}
K.~Hinterbichler, ``{Theoretical Aspects of Massive Gravity},''
\href{http://arxiv.org/abs/1105.3735}{{\ttfamily arXiv:1105.3735 [hep-th]}}.

\bibitem{Goldhaber:2008xy}
A.~S. Goldhaber and M.~M. Nieto, ``{Photon and Graviton Mass Limits},''
  \href{http://dx.doi.org/10.1103/RevModPhys.82.939}{{\em Rev.Mod.Phys.}
  {\bfseries 82} (2010) 939--979},
\href{http://arxiv.org/abs/0809.1003}{{\ttfamily arXiv:0809.1003 [hep-ph]}}.

\bibitem{ArkaniHamed:2002sp}
N.~Arkani-Hamed, H.~Georgi, and M.~D. Schwartz, ``{Effective field theory for
  massive gravitons and gravity in theory space},'' {\em Ann. Phys.} {\bfseries
  305} (2003) 96--118,
\href{http://arxiv.org/abs/hep-th/0210184}{{\ttfamily arXiv:hep-th/0210184}}.

\bibitem{deRham:2010gu}
C.~de~Rham and G.~Gabadadze, ``{Selftuned Massive Spin-2},''
  \href{http://dx.doi.org/10.1016/j.physletb.2010.08.043}{{\em Phys.Lett.}
  {\bfseries B693} (2010) 334--338},
\href{http://arxiv.org/abs/1006.4367}{{\ttfamily arXiv:1006.4367 [hep-th]}}.

\bibitem{Deffayet:2010zh}
C.~Deffayet, S.~Deser, and G.~Esposito-Farese, ``{Arbitrary $p$-form
  Galileons},'' \href{http://dx.doi.org/10.1103/PhysRevD.82.061501}{{\em
  Phys.Rev.} {\bfseries D82} (2010) 061501},
\href{http://arxiv.org/abs/1007.5278}{{\ttfamily arXiv:1007.5278 [gr-qc]}}.

\bibitem{Padilla:2010de}
A.~Padilla, P.~M. Saffin, and S.-Y. Zhou, ``{Bi-galileon theory I: Motivation
  and formulation},'' {\em JHEP} {\bfseries 1012} (2010) 031,
\href{http://arxiv.org/abs/1007.5424}{{\ttfamily arXiv:1007.5424 [hep-th]}}.

\bibitem{Hinterbichler:2010xn}
K.~Hinterbichler, M.~Trodden, and D.~Wesley, ``{Multi-field galileons and
  higher co-dimension branes},''
  \href{http://dx.doi.org/10.1103/PhysRevD.82.124018}{{\em Phys.Rev.}
  {\bfseries D82} (2010) 124018},
\href{http://arxiv.org/abs/1008.1305}{{\ttfamily arXiv:1008.1305 [hep-th]}}.

\bibitem{D'Amico:2012zv}
G.~D'Amico, G.~Gabadadze, L.~Hui, and D.~Pirtskhalava, ``{Quasi-Dilaton: Theory
  and Cosmology},''
\href{http://arxiv.org/abs/1206.4253}{{\ttfamily arXiv:1206.4253 [hep-th]}}.

\bibitem{Boulware:1973my}
D.~G. Boulware and S.~Deser, ``{Can gravitation have a finite range?},''
\href{http://dx.doi.org/10.1103/PhysRevD.6.3368}{{\em Phys. Rev.} {\bfseries
  D6} (1972) 3368--3382}.

\bibitem{Lanczos:1938sf}
C.~Lanczos, ``{A Remarkable property of the Riemann-Christoffel tensor in four
  dimensions},''
{\em Annals Math.} {\bfseries 39} (1938) 842--850.

\bibitem{Lovelock:1971yv}
D.~Lovelock, ``{The Einstein tensor and its generalizations},''
\href{http://dx.doi.org/10.1063/1.1665613}{{\em J.Math.Phys.} {\bfseries 12}
  (1971) 498--501}.

\bibitem{Myers:1987yn}
R.~C. Myers, ``{Higher Derivative Gravity, Surface Terms and String Theory},''
\href{http://dx.doi.org/10.1103/PhysRevD.36.392}{{\em Phys.Rev.} {\bfseries
  D36} (1987) 392}.

\bibitem{Charmousis:2005ey}
C.~Charmousis and R.~Zegers, ``{Matching conditions for a brane of arbitrary
  codimension},'' \href{http://dx.doi.org/10.1088/1126-6708/2005/08/075}{{\em
  JHEP} {\bfseries 0508} (2005) 075},
\href{http://arxiv.org/abs/hep-th/0502170}{{\ttfamily arXiv:hep-th/0502170
  [hep-th]}}.

\bibitem{Charmousis:2005ez}
C.~Charmousis and R.~Zegers, ``{Einstein gravity on an even codimension
  brane},'' \href{http://dx.doi.org/10.1103/PhysRevD.72.064005}{{\em Phys.Rev.}
  {\bfseries D72} (2005) 064005},
\href{http://arxiv.org/abs/hep-th/0502171}{{\ttfamily arXiv:hep-th/0502171
  [hep-th]}}.

\bibitem{Hassan:2011zd}
S.~Hassan and R.~A. Rosen, ``{Bimetric Gravity from Ghost-free Massive
  Gravity},'' \href{http://dx.doi.org/10.1007/JHEP02(2012)126}{{\em JHEP}
  {\bfseries 1202} (2012) 126},
\href{http://arxiv.org/abs/1109.3515}{{\ttfamily arXiv:1109.3515 [hep-th]}}.

\bibitem{Hassan:2011hr}
S.~Hassan and R.~A. Rosen, ``{Resolving the Ghost Problem in non-Linear Massive
  Gravity},'' \href{http://dx.doi.org/10.1103/PhysRevLett.108.041101}{{\em
  Phys.Rev.Lett.} {\bfseries 108} (2012) 041101},
\href{http://arxiv.org/abs/1106.3344}{{\ttfamily arXiv:1106.3344 [hep-th]}}.

\bibitem{Hassan:2011tf}
S.~Hassan, R.~A. Rosen, and A.~Schmidt-May, ``{Ghost-free Massive Gravity with
  a General Reference Metric},''
  \href{http://dx.doi.org/10.1007/JHEP02(2012)026}{{\em JHEP} {\bfseries 1202}
  (2012) 026},
\href{http://arxiv.org/abs/1109.3230}{{\ttfamily arXiv:1109.3230 [hep-th]}}.

\bibitem{Hassan:2011ea}
S.~Hassan and R.~A. Rosen, ``{Confirmation of the Secondary Constraint and
  Absence of Ghost in Massive Gravity and Bimetric Gravity},''
\href{http://arxiv.org/abs/1111.2070}{{\ttfamily arXiv:1111.2070 [hep-th]}}.

\bibitem{deRham:2011qq}
C.~de~Rham, G.~Gabadadze, and A.~J. Tolley, ``{Helicity Decomposition of
  Ghost-free Massive Gravity},''
  \href{http://dx.doi.org/10.1007/JHEP11(2011)093}{{\em JHEP} {\bfseries 1111}
  (2011) 093},
\href{http://arxiv.org/abs/1108.4521}{{\ttfamily arXiv:1108.4521 [hep-th]}}.

\bibitem{deRham:2011rn}
C.~de~Rham, G.~Gabadadze, and A.~Tolley, ``{Ghost free Massive Gravity in the
  St\'uckelberg language},''
\href{http://arxiv.org/abs/1107.3820}{{\ttfamily arXiv:1107.3820 [hep-th]}}.

\bibitem{Mirbabayi:2011aa}
M.~Mirbabayi, ``{A Proof Of Ghost Freedom In de Rham-Gabadadze-Tolley Massive
  Gravity},''
\href{http://arxiv.org/abs/1112.1435}{{\ttfamily arXiv:1112.1435 [hep-th]}}.

\bibitem{Hinterbichler:2012cn}
K.~Hinterbichler and R.~A. Rosen, ``{Interacting Spin-2 Fields},''
\href{http://arxiv.org/abs/1203.5783}{{\ttfamily arXiv:1203.5783 [hep-th]}}.

\bibitem{Siegel:1993sk}
W.~Siegel, ``{Hidden gravity in open string field theory},''
  \href{http://dx.doi.org/10.1103/PhysRevD.49.4144}{{\em Phys.Rev.} {\bfseries
  D49} (1994) 4144--4153},
\href{http://arxiv.org/abs/hep-th/9312117}{{\ttfamily arXiv:hep-th/9312117
  [hep-th]}}.

\bibitem{Goon:2010xh}
G.~L. Goon, K.~Hinterbichler, and M.~Trodden, ``{Stability and superluminality
  of spherical DBI galileon solutions},''
  \href{http://dx.doi.org/10.1103/PhysRevD.83.085015}{{\em Phys.Rev.}
  {\bfseries D83} (2011) 085015},
\href{http://arxiv.org/abs/1008.4580}{{\ttfamily arXiv:1008.4580 [hep-th]}}.

\bibitem{Hassan:2011vm}
S.~Hassan and R.~A. Rosen, ``{On Non-Linear Actions for Massive Gravity},''
  \href{http://dx.doi.org/10.1007/JHEP07(2011)009}{{\em JHEP} {\bfseries 1107}
  (2011) 009},
\href{http://arxiv.org/abs/1103.6055}{{\ttfamily arXiv:1103.6055 [hep-th]}}.

\bibitem{Ostrogradski}
M.~Ostrogradski {\em Mem. Ac. St. Petersbourg} 385.

\bibitem{Koyama:2011wx}
K.~Koyama, G.~Niz, and G.~Tasinato, ``{The Self-Accelerating Universe with
  Vectors in Massive Gravity},''
  \href{http://dx.doi.org/10.1007/JHEP12(2011)065}{{\em JHEP} {\bfseries 1112}
  (2011) 065},
\href{http://arxiv.org/abs/1110.2618}{{\ttfamily arXiv:1110.2618 [hep-th]}}.

\end{thebibliography}\endgroup

\end{document}